\documentclass{aa}  
\usepackage{graphicx}
\usepackage{subfigure}

\usepackage{txfonts}
\begin{document}
   \title{ A Search For Supernova Remnants in The Nearby Spiral Galaxy M74 (NGC 628) }

    \author{E. Sonba\c{s}
          \inst{1}
           \and
           A. Aky\"uz \inst{2}
          \and
          \c{S}. Balman \inst{3}
         \and
         M.E. \"Ozel \inst{4}
        }
      
   \offprints{E. Sonba\c{s}}

   \institute{University of Ad{\i}yaman, Department of Physics, 02040 Ad{\i}yaman, Turkey 
             \and
	     University of \c{C}ukurova, Department of Physics, 01330 Adana, Turkey
	     \and
              Dept. of Physics, Middle East Technical University, 06531 Ankara, Turkey 
            \and
            \c{C}a\u{g} University, Faculty of Arts and Sciences, 33800 Tarsus, Turkey   
            }
 \date{Received -; accepted -}


\abstract
   {}
   { An optical search is carried out for supernova remnants (SNRs) in the Sc type nearby spiral galaxy M74, using ground-based observations at  
        TUBITAK National Observatory (TUG, Antalya/Turkey) and Special Astrophysics Observatory (SAO, Russia). Observations are supplemented by the spectral analysis of archived X-ray data from $XMM$-$Newton$\ and $Chandra$.}
   { A survey of M74 covering $\sim$ 9 arcmin$^{2}$ with [S II], H$\alpha$, and their continuum filters. Interference filter images of M74 are obtained 
with 1.5 m Russian Turkish Telescope (RTT150) at TUG and  spectral data are taken with the 6 m Bolsoi Azimuthal Telescope (BTA) at SAO. 
The emission nebulae with continuum-subtracted line ratio values of [S II]$\lambda$$\lambda$6716,6731 /H$\alpha$ $\geq$ 0.4 are 
identified as SNRs. A follow-up spectroscopy is obtained to confirm optical SNR identifications.}
   { We have identified nine new SNR candidates in  M74 with [S II]/H$\alpha$ $\geq$ 0.4 as the basic criterion. We obtain 
[S II]/H$\alpha$ ratio in the range from 0.40 to 0.91 and H$\alpha$ intensities from 2.8 $\times$ $10^{-15}$ erg cm$^{-2}$ s$^{-1}$ to 
1.7 $\times$ $10^{-14}$ erg cm$^{-2}$ s$^{-1}$. We also present spectral follow-up observations of these  SNR candidates. However, we are able 
	to spectrally confirm only three  of them (SNR2, SNR3, and SNR5). The 
lack of confirmation for the rest 
 might be due to the contamination by the nearby H II emission regions as well as due to the inaccurate positioning of the long slit on these objects. 
In addition, we search the $XMM$-$Newton$\ and $Chandra$ Observatory archival data for the X-ray counterparts to the optically identified candidates. We find positional coincidence with only three SNR candidates, SNR1, SNR2, and SNR8. 
The spectrum of SNR2 yields a shock temperature of 10.8 keV with an ionization timescale of 1.6 $\times$ 10$^{10}$ s cm$^{-3}$
indicating a relatively young remnant in an early Sedov phase which is not supported by our optical wavelength
analysis. Given the high luminosity of 10$^{39}$ erg s$^{-1}$ and the characteristics of the X-ray 
spectrum, we favor an  Ultra Luminous X-ray Source interpretation for this source associated with an SNR. 
We calculate  an X-ray flux upper limit of 9.0 $\times$ $10^{-15}$ erg cm$^{-2}$ s$^{-1}$ for the rest of the SNRs including
spectroscopically identified SNR3 and SNR5.}
   {}


   \maketitle
%

\section{Introduction}

M74 (NGC 628) is an Sc type spiral galaxy with a $\sim$ 6$^{\circ}$ inclination angle at an assumed distance of 7.3 Mpc 
(Sharina et al.1996, Soria et al. 2004). It is the brightest member of the small M74 group in the Pisces constellation.  M74 has 
spiral arms well formed with bright blue star clusters and dark cosmic dust lanes. In the early 2000, 
two supernovae (SN 2002ap and SN 2003gd) were identified in M74 and studied extensively. 
The supernova event SN 2002ap is one of the four 
Type Ic SNe observed in X-rays (Van Dyk et al. 2003). 
SN 2003gd, which is found to be a nearby Type II-P (plateau) event 
has a confirmed red giant progenitor (Hendry et al. 2005).  

 Leievre \& Roy (2000) have recovered well over 100 small and isolated H II regions and measured their fluxes in the extreme outer disk of M74 with 
R$>$R$_{25}$  
making use of deep H$\alpha$ narrowband imaging. They show that massive star formation rate measured by the azimuthally averaged H$\alpha$  
surface brightness decreases monotonically from the center out to R$\sim$20 kpc beyond which it drops rapidly.

 Elmegreen et al. (2006) have examined the size and luminosity distributions of the star forming regions in the galaxy with the Hubble Space 
Telescope. Their results suggest clumping of stars having a scale free nature in the galaxy disk. Fathi et al. (2007) have studied the internal kinematics of M74 
over its entire face using Fabry-Perot interferometry with a good angular resolution, and  
confirmed the presence of an inner rapidly rotating disc-like component 
caused by the slow secular evolution of the large-scale spiral arms together with the oval 
structure. They also detect over 300 H II regions in the galaxy with calibrated luminosities and diameters up to 
about 300 pc using continuum subtracted narrow band images in H$\alpha$.

M74 is observed in the X-rays using the $Chandra$ and $XMM$-$Newton$\ data by Soria et al. (2004). 74 discrete X-ray sources are detected 
in the combined $Chandra$ observations 
to a detection limit of $\approx$ 6 $\times$ 10$^{36}$ erg s$^{-1}$. 
They estimate 15$\%$ of the M74 discrete sources to have soft
colors. They also calculate thermal SNRs with luminosities $\approx$ 2 $\times$ 10$^{37}$ erg s$^{-1}$. 
Two bright  X-ray sources (XMMU J013636.5+155036, CXOU J13651.1+154547) are detected by $XMM$-$Newton$\ in M74 
(Soria et al. 2004), while 
only the latter is detected by Chandra (Krauss et al. 2003). It is established that both of the sources are
variable. It is also argued that they can be Ultra Luminous X-ray sources (ULX), however their true nature 
remains unclear.

   Radio observations of M74 have been carried out during the Arecibo Galaxy Environment Survey (AGES). 
This survey is designed to investigate M74  group environment in 21 cm with a better sensitivity and 
higher spatial and velocity resolution
 than previous observations (Auld et al. 2006). They obtain the spatial distribution of HI for M74 and other 
selected galaxies.

 In the UV regime, the surface brightness and color profiles for M74  are calculated by Cornett et al. (1994). 
They show that M74 disk has sustained significant star formation over the last five hundred million years.

  In this paper, we present our results on the search for new supernova remnant (SNR) candidates 
in M74 using the observations by RTT150 and 6 m-BTA 
telescopes. 
We also make use of archived $Chandra$ data to look for their X-ray 
counterparts. SNR studies are important in the theories of interstellar medium (ISM),
and star formation 
since supernovae inject large amounts of matter, and energy into the 
ISM. However, in spite of the quite 
large number of Galactic SNRs their observations are impeded by several limitations such as the
uncertainty in distances to individual objects and high extinctions along the 
line of sight in several regions of the Galactic plane. The limitations and uncertainties are inherently much 
less in extragalactic samples. Assuming that all SNRs are of the same 
distance to us for a given galaxy, we can easily compare  their 
observed properties. 
Relative positions of such SNRs are determined with more precision. 
Once we know the positions of SNRs, their distances with respect to H II 
regions and spiral arms can easily be calculated (Matonick \& Fesen 1997; 
Blair \& Long 1997, 2004).

SNRs are identified in a number of nearby spiral galaxies using optical observations 
(e.g. D$^\prime$Odorico, Dopita \& Benveuti 1980; Braun \& Walterbos 1993; 
Magnier et al. 1995; Matonick \& Fesen 1997; Matonick et al. 1997; Gordon et al. 1998,1999; Blair \& Long 1997, 2004; Sonbas et al. 2009), 
X-ray observations
 (Pence et al. 2001; Ghavamian et al. 2005), and radio observations (Lacey et al. 1997; Lacey \& Duric 2001; Hyman et al. 2001). 
Multiwavelength surveys of SNRs are also carried out by Pannuti et al. (2000, 2002, 2007). 

In our work, we use a well known and accepted criterion (S II/H$\alpha$ $\geq$ 0.4) proposed by Mathewson \& Clarke (1973) to differentiate SNRs from 
typical H II regions. 
 Rationale for this lies in the fact that in a typical H II region, sulfur is usually expected to be in the form of $S^{++}$ due to strong photoionizing 
fluxes from central 
hot stars,  making the ratio [S II]/H$\alpha$, typically, in the range $\sim$ 0.1 - 0.3. Shock waves produced by  SN explosion
 propagate through the surrounding medium. The matter
cools sufficiently behind these waves and variety of ionization states occur including S$^{+}$. This might be the reason for 
the increased [S II]/H$\alpha$ ratio observed in SNRs.  It follows that almost all discrete emission nebulae satisfying the above criterion are expected
 to be shock-heated.

The organization of the paper is as follows; In section 2, we  
discuss our imaging and spectroscopic observations and related data reduction. 
Search results in the optical band, the identification of SNRs, 
and search for their X-ray counterparts are presented in section 3. Finally, conclusions and discussions of our results 
are provided in section 4. 

\section{Observations and data reduction}
     
     \subsection{Imaging}

The imaging observations of M74 were performed with the 1.5 m Russian Turkish Telescope (RTT150) at TUBITAK National Observatory (TUG) in Turkey.
 M74 images were obtained using a Cassegrain imaging CCD Loral Lick 2k $\times$ 2k back-illuminated and anti-reflection coated (single layer, 700 
\AA\ of hafnium oxide, HfO2), 
2048 $\times$ 2048 pixel CCD with a plate scale of 0$\arcsec$.26 pixel$^{-1}$, giving a 9$\arcmin$.1 $\times$ 9$\arcmin$.1 field of view (FOV). We used narrowband interference filters 
centered on the lines of [S II], H$\alpha$, and continuum filters to remove starlight from H$\alpha$ and [S II] images. The interference filter characteristics used in these observations 
are shown in Table 1. Only one field image for each filter was obtained for M74 on September 19 and 20, 2004. We have exposure times of 12 $\times$ 300 sec for the [S II], H$\alpha$ filters, and 4 $\times$ 300 sec for the two continuum filters. This way, we were able to obtain deeper field images with higher signal-to-noise ratio for the faintest objects. 

The European Southern Observatory Munich Image Data Analysis System (ESO-MIDAS) software environment (version 08FEBpl1.1) was 
used to reduce the data (http://www.eso.org). We observed two or three standard stars each night from the list of Oke (1974) and Stone (1977) 
to determine
 the average flux conversion factors (FCF).  We used the technique from Jacoby et al. (1987) to calculate this factor in the units of
 erg\  cm$^{-2}$\ s$^{-1}$/\ ADU s$^{-1}$. FCF allows us to convert measured counts from an image to a flux in units of 
erg\ cm$^{-2}$\ s$^{-1}$. 
Bias and flat-field frames for each image were also obtained. This way each exposure was bias subtracted, 
trimmed and flat-fielded. The cosmic hits 
were also removed from each [S II] and H$\alpha$ image. Astrometry was applied to the resulting
 [S II] and H$\alpha$ images to include 
world coordinate system information into the FITS header of the individual images of M74. 
Red stars from USNO A2.0 catalog (Monet et al. 1998)
 were used to check positions of the stars. In Figure 1, we show the positions of SNR candidates 
overlaid on the Digitized Sky Survey (DSS)
 image of the M74 galaxy.

\subsection{Spectral data} 

In order to resolve the [S II]$\lambda\lambda$6716,6731 lines and to separate [NII]$\lambda\lambda$6548,6583 lines from 
H$\alpha$ providing a more accurate [S II]/H$\alpha$ values than imaging ratio values, we carried out spectral follow-up observations.
The spectral data of our SNR candidates were obtained by SCORPIO
(The Spectral Camera with the Optical Reducer for Photometrical and Interferometrical Observations) which  
was mounted to 6m BTA with a 2048 $\times$ 2048 pixel CCD (Afanasiev \& Moiseev 2005). 
A 3500-7200 \AA\ spectral coverage at 10 \AA\ resolution and a 1$\arcsec$ wide slit were used during the observations.

The basic data reductions, flux and wavelength calibrations and interstellar extinction correction were performed using IDL 
codes and IRAF (Image Reduction and Analyses Facility) packages (http://iraf.noao.edu/docs/docmain.html). 
Spectrophotometric standard stars from Oke (1974) and Stone (1977) 
catalogs were also observed each night. We derived fluxes from the spectrophotometric standards and used these to 
calibrate the fluxes for spectral lines
 in our SNR spectra. Biases, internal lamp flats, and Ne-Ar calibration lamp frames were obtained for each 
observation set. The $\textit{splot}$ IRAF routine was used to measure the emission line fluxes of [S II] and H$\alpha$. While obtaining spectral data of SNR candidates, a bright star was placed on the slit along with the target object. Galactic extinction correction was applied to the SNR spectra as part of the spectral analysis (Cardelli et al. 1989).  

\section{Results and discussion}

\subsection{Results from Optical Imaging Analysis}

We used the technique described by Blair \& Long (1997) in order to identify the SNR candidates. 
The identification technique was constructed through the comparison of continuum - subtracted 
[S II]$\lambda\lambda$6716,6731 and H$\alpha$ images. Eventually, emission nebulae that have 
region-integrated values of [S II]/H$\alpha$$\geq$ 0.4 were identified as the SNR candidates. 
Preliminary search to find these candidates were carried out through a comparison of the continuum 
subtracted [S II] and the H$\alpha$ subfield images by a blinking technique. In the process, a full-field 
image of the galaxy was divided into regions of about 2$\arcmin$ by 2$\arcmin$ squares for the visual 
inspection and assessment of the fields for the candidates. 

When a point like source satisfied the criterion [S II]/H$\alpha$ $\geq$ 0.4 we then marked it as an SNR 
candidate. After the assessment of all 2$\arcmin$ by 2$\arcmin$ subfields of M74, continuum subtracted 
images were used to determine the ratio of [S II]/H$\alpha$ counts. During this process, background 
counts were subtracted from the [S II]/H$\alpha$ ratio images using annuli farther away, but centered
on the selected candidates. The background extraction areas were also normalized to the source
extraction areas.
At the assumed distance of 7.3 Mpc for M74, our aperture size corresponds to $\sim$ 50 pc. 
During our interference filter imaging observations seeing was 1.4$\arcsec$, thus we can not extract 
any radius information for our SNR candidates since it was limited with the seeing for the night.

We found nine emission nebulae consistent with  [S II]/H$\alpha$ $\geq$ 0.4 criterion
for SNR identification with the technique described above. The list of new SNR candidates 
with corrected flux ratios are given in Table 2. H$\alpha$ fluxes for SNR2, SNR3, and SNR5 are
taken from spectral measurements while remaining ones are from image data.
A limiting flux sensitivity of $\sim$
10$^{-15}$  erg cm$^{-2}$s$^{-1}$ was calculated 
by choosing a structure with a limiting magnitude in the field of the galaxy.
M74 covers a total area of about 10$\arcmin$.5 $\times$ 9$\arcmin$.5 in the sky. Since FOV  
of our CCD is 9$\arcmin$.1 $\times$ 9$\arcmin$.1,  each filter image covered
almost the full face of M74. Therefore only a single observation for each filter was required. We zoomed on the
full continuum-subtracted H$\alpha$ image of M74,  and extracted four subfields, each of size
3$\arcmin$.5 $\times$ 2$\arcmin$.5, and marked our SNR candidates on these images as displayed in Figure 2. 

\subsection{The Follow-up Optical Spectroscopy}

We also conducted spectral observations to confirm the optical SNR candidates detected by optical imaging analysis. 
Among the nine candidates detected in M74, we were able to confirm only three of them (SNR2, SNR3, and SNR5) with the derived 
specific line ratios  [S II]/H$\alpha$ of 0.46, 0.91, and 0.58 respectively. Their optical 
spectra are shown in Figures 3-5. Observed line intensities relative to H$\beta$, the E$_{(B-V)}$ 
values and  H$\alpha$ intensities  for the spectra of these three SNRs are given in Table 3. 
The rest of the six SNR candidates (SNR1, SNR4, SNR6, SNR7, SNR8, SNR9) could not be confirmed using the
spectroscopic line ratio values. Several factors could be affecting the spectral [S II]/H$\alpha$ line ratio values. 
For instance, the long slit used for light acquisition may not always be accurately pointed on the objects. In such cases  
contamination from nearby or overlapping H II emission regions could be affecting the line 
ratios. This seems likely for SNR1, SNR6, SNR7, SNR8, and SNR9 that were quite close to extensive H II 
emission regions as shown in Figure 6. The spatial coordinates and sizes of these H II regions were
 taken from the catalog of Fathi et al. (2007) (The catalog can also be reached at http://cdsweb.u-strasbg.fr/cgi-bin/qcat?J/A+A/466/905/). 
 We thought that there could be significant H II emission contamination into these sources, as was foreseen by Matonick \& Fesen (1997), and Blair \& Long (2004). They discuss some of their SNR candidates in
relation to closeby  H II regions (e.g., their SNR5 in NGC6946, SNR39 in M101, and SNR19 in M81) with the angular separations of the SNRs being up to 15$\arcsec$ - 40$\arcsec$ from H II regions. The
angular separations of our candidates SNR1, SNR6, SNR7, SNR8, and SNR9 from the nearby H II regions calculated using our images
were found in the range  5$\arcsec$ to 10$\arcsec$ in M74. This corresponds to approximately the same distance scale when the relative distances of target galaxies are considered. On the other hand, we estimate that SNR1, SNR4, SNR8, SNR9 have significant 
total [O I] emissions reaching about from 20$\%$ to 60$\%$ of the H$\beta$ intensity. This is considered
 to be a secondary shock-heated gas indicator. In turn, presence of a detectable [O I]$\lambda$6300 emission 
is considered to be  a viable argument for  SNR identifications, in general (Blair \& Long 2004).

We note that, the detected SNR distribution in M74 appears mostly in the southern field of the galaxy. 
Similar asymmetric distributions are observed in some other spiral galaxies such as NGC 2403 and NGC 2903 
(Matonick et al. 1997; Sonbas et al. 2009).  Presently, the leading explanation
for such asymmetries is that in the half of a given galaxy where H II regions are brighter, less SNRs are detected.
This seems to be the case also for M74. We believe that this is not an artifact of our imaging or SNR identification techniques. We propose that, such asymmetries need to be discussed further with more galaxy examples and observations.

At the most basic level, average electron density in a nebula is usually measured 
through effects of collisional de-excitation. This is usually done by the comparison of 
intensities of two lines from the same ion, emitted by different levels with the same excitation energy. 
One of the best examples of this procedure makes use of the electron density ratios [S II]$\lambda$6716/[S II]$\lambda$6731 and [OII]$\lambda$3729/[OII]$\lambda$3726. 
 In our work we used [S II]$\lambda$6716/[S II]$\lambda$6731 line ratio from SNR spectra to calculate the electron density, N$_e$ .
 The Space Telescope Science Data Analysis System (STSDAS) task $nebular.temden$, based on the five-level atom approximation, 
calculates the required electron density given the  
electron temperature. Our calculated $N_e$ value is 190 $\pm$ 100 cm$^{-3}$ for SNR3 corresponding to the line ratio of 1.24 for an assumed
 electron temperature of T = 10$^{4}$ K. 
The reason for the error term quoted here arises from the fact that the theoretical model loses its sensitivity for electron densities $<$ 100 cm$^{-3}$ which is in line with our spectral measurement sensitivities.  
With the same reasoning, although the line ratio values of [S II]$\lambda$6716/[S II]$\lambda$6731 $>$ 1.46 is considered
to correspond to a low density limit with $N_e$ $\leq$ 10 cm$^{-3}$ (Osterbrock 1989), we can only quote an electron density for both SNR2 and SNR5 of simply $<$ 100 cm$^{-3}$. 
Using the shock abundance models by Dopita et al. (1984), the elemental abundances can be determined 
from the SNR spectra. Their models assume a shock velocity around or less than 100 km s$^{-1}$ which is considered
a low shock velocity. Same authors also make use of the fact that the behavior of [OIII] line is the most sensitive
indicator of shock velocity variations. These variations fall rapidly for values less than 85 km s$^{-1}$.   
In our case, [OIII]$\lambda$5007/H$\beta$ emission line ratios for SNR2, SNR3 and SNR5 were found as 0.27, 1.22 and 
1.6, respectively, corresponding to nebular propagation velocities of $<$ 85 km s$^{-1}$ (see Figure 5  of Dopita et al. 1984). There are many SNRs which have low shock velocity
values in a number of nearby galaxies (Matonick \& Fesen 1997; Blair \& Long 2004).

Among the working methods to separate SNRs from H II regions are the plots of [S II]/H$\alpha$ 
versus H$\beta$ flux and  [S II]/H$\alpha$ versus [NII]/H$\alpha$ line ratios.
 It was noted that in a number of nearby galaxies many SNR candidates show enhanced [NII] 
and very weak H$\beta$ emissions when compared with the H II regions.
We adopted the method
by Blair \& Long (2004) based on their Figure 18 where the
same two plots for SNRs in M83  are given. 
When we inserted our ratio values for detected SNRs into these plots (Figure 7), we find similar tendencies.
Our new SNR candidates  are also located in the lower end as
would be expected. The observed low H$\beta$ flux values are indicative of faint ionized
hydrogen regions.

\subsection{Comparison with X-ray Observations} 

In a  cross-check with the X-ray point source list of M74, three X-ray sources 
(CXOUJ013631.7+154848, CXOUJ013644.0+154908, CXOUJ013646.0+154422, Soria et al. 2004) 
are found coincident with our optically detected SNRs (SNR1, SNR2, SNR8) taking $\sim$ 8$\arcsec$
positional error circle around the objects. The error in association of SNR1 and SNR8 is $\sim$ 5$\arcsec$
and SNR2 is $\sim$ 3$\arcsec$ .
In Table 4 we list the optical and X-ray associations for these SNRs.
Among these, only SNR2 is a spectrally confirmed candidate. We used an archival Chandra pointed  
observation of M74  obtained in 2001 June 19 (with an  exposure 
of 46.3 ksec) to derive a spectrum of the SNR. We performed standard
point source spectral analysis on the data and extracted source photons from the location of SNR2 
with a circular extraction region of 12$\arcsec$.5 and background photons were extracted from
a region normalized to the source photon extraction region. Entire analysis was performed
using CIAO 3.4 (http://asc.harvard.edu/ciao3.4/index.html) and XSPEC 12.5.0ac software (Arnaud 1996). 

We measured a count rate of 
(1.5$\pm$0.5) $\times$ 10$^{-3}$ cts s$^{-1}$ for  SNR2. We derived a source 
spectrum and fitted this spectrum using the XSPEC model PSHOCK, a
non-equilibrium ionization plasma emission model for shock-heated plasma
appropriate for SNRs (Borkowski et al. 1996). 
The final Chandra spectrum of SNR2 is displayed in Figure 8. 
The resulting spectral parameters are 
N$_{H}$=1.1$^{+0.4}_{>}$ $\times$ 10$^{22}$ cm$^{-2}$, 
kT$_{plasma}$=10.8$^{+26.0}_{-9.4}$ keV,
the ionization timescale $\tau$ =n$_{0}$t=1.6$^{+9.5}_{>}$ $\times$ 10$^{10}$ s cm$^{-3}$,
and the model normalization K=1.1$^{+2.8}_{>}$ $\times$ 10$^{-5}$ cm$^{-5}$; the lower limit of the normalization is unconstrained. 
The $\chi^2_{\nu}$ of the fit is 1.3.
The unabsorbed source flux is about 1.7 $\times$ 10$^{-13}$ erg cm$^{-2}$ s$^{-1}$ which
translates to an X-ray luminosity $\sim$ 1.1 $\times$ 10$^{39}$ erg s$^{-1}$ 
at the source distance of 7.3 Mpc. 
The derived plasma temperature yields a shock speed of 3100 km s$^{-1}$ 
($kT_s=(3/16)\mu m_H (v_s)^2$) using Rankine-Huguenot
jump conditions. The model 
normalization can be used to calculate the electron density of the 
shocked material as n$_{e}$=0.45 cm$^{-3}$ (filling factor f=1) 
given a volume limited by the seeing that corresponds to 50
pc in size at the source distance. This value is consistent with the
density derived from our optical line ratio analysis n$_{e}$ $<$ 10 cm$^{-3}$.
Assuming that the ambient density n$_{0}$ is not going 
to be more than the electron density, the ionization timescale yields a 
minimum life time of about 1120 years for SNR2. This indicates that SNR2 
maybe in the  Sedov phase of its evolution. 

It is important to note that the calculated luminosity for SNR2 is large for
an SNR in the Sedov phase of evolution and the calculated expansion speed can not be
supported with the blue shifts derived from the optical spectroscopy.
Given this result, we also suspect that SNR2 may be a ULX (e.g. XMMU J013636.5+155036; Soria $\&$ Kong
2002). We modeled the spectrum of SNR2 with a power-law yielding an
N$_{H}$=2.6$^{+0.5}_{>}$ $\times$ 10$^{21}$ cm$^{-2}$,
a $\Gamma$=1.8$^{+1.5}_{-0.9}$ keV  and a normalization 
K$_{pl}$=3.7$^{+0.9}_{>}$ $\times$ 10$^{-6}$
with a  $\chi^2_{\nu}$=1.2. Adding a blackbody component along with the
power-law gives a kT $\sim$ 0.6-2 keV and steepens the power-law index to 1.8-5.4. A single
blackbody fit gives a large  $\chi^2_{\nu}$ above 1.5. 

Given the high shock plasma
temperature yielding expansion speed that can not be supported with the optical data and that
the X-ray luminosity is of the order of 10$^{39}$ erg s$^{-1}$, we do not favor an old SNR interpretation
for SNR2 as expected from the optical data analysis. We favor a ULX interpretation associated with
a SNR. We expect that the very slow (less than 85 km s$^{-1}$) expansion speed from the
interpretation of the optical data will not yield detectable X-ray emission at the distance of
M74. Then, any emission resulting from the point source remnant of the SN explosion, will dominate
the X-ray emission, though the SNR may dominate the optical emission. 
The ULX is expected to rejuvenate and ionize the nebulosity around it.
Associations between
SNRs and ULXs have been found or suggested before (Roberts et al. 2003, Abolmasov et al. 2008, 
Di Stefano $\&$ Kong 2004; Di Stefano, et al. 2010).	

We also checked the other two sources, SNR1 and SNR8.
These sources were detected with (1.0$\pm$0.5) $\times$ 10$^{-3}$ cts s$^{-1}$ and 
(2.7$\pm$0.5) $\times$ 10$^{-3}$ cts s$^{-1}$, respectively. 
The fits with the PSHOCK model the two sources result in
a neutral hydrogen absorption of about 0.9 $\times$ 10$^{22}$ cm$^{-2}$, and a kT$_{plasma}$ of about
34-73 keV with an ionization timescale $\tau$ in a range 2-5 $\times$ 10$^{10}$ s cm$^{-3}$. 
The spectral parameters yield an unabsorbed X-ray flux of 5 $\times$ 10$^{-15}$ erg cm$^{-2}$ s$^{-1}$
and 1 $\times$ 10$^{-13}$ erg cm$^{-2}$ s$^{-1}$ for SNR1 and SNR8, respectively, translating to
an X-ray luminosity of $\sim$ 5 $\times$ 10$^{37}$ erg s$^{-1}$ and $\sim$ 7 $\times$ 10$^{38}$ erg s$^{-1}$.
Given  these spectral parameters and assuming these are SNRs we can speculate that 
the expansion speed is 2-3 times larger than SNR2
for these two SNRs, the electron density is about 1.4 to 2.3 times less than SNR2, 
and
the minimum life time is about 1570-2600 years for SNR1 and SNR8 compared with SNR2. 
In general, one expects that the 
SNR shock decelerates in time, thus these longer life times are in contradiction with
the faster shock speeds and high X-ray temperatures
detected for these two SNRs in comparison with SNR2. These shock temperatures and expansion
speeds are not supported by our optical data. 
We note that these two sources are not confirmed SNRs with our ratio measurements using optical 
spectroscopy. Thus, we 
suspect them as SNRs, we favor an X-ray binary or an AGN interpretation for the two sources. For the same three sources $XMM$-$Newton$\ data yielded confirming spectra.
Therefore, no details are repeated here. 

Finally, we calculate an X-ray flux limit of 
$<$ 9.0 $\times$ 10$^{-15}$ erg cm$^{-2}$ s$^{-1}$ with a luminosity limit of $<$ 5.0 $\times$ 10$^{37}$ erg s$^{-1}$ for the
rest of the SNRs in our list including the two other spectrally confirmed sources SNR3 and SNR5 since
we find no positional coincidence with any of the source listings of M74.

\section{Conclusions}

We conducted a survey of SNRs using optical imaging and spectroscopic measurements in 
the nearby spiral galaxy M74. In this survey, we used blinking between continuum-subtracted 
H$\alpha$ and continuum-subtracted [S II] images to deduce SNR candidates for a given ratio between the 
lines.  The SNRs were confirmed by additional spectroscopic 
observations and finally, comparing the optically detected SNRs with archived Chandra observations. 
Nine SNRs were identified in M74 by our method. Three of these have shown positional 
coincidences with X-ray sources. SNR2 is our best candidate to be an SNR
using optical wavelength analysis and the X-ray source is consistent with a ULX interpretation,
making it one of the few associations between a SNR and a ULX.
We calculated an X-ray flux upper limit of 9.0 $\times$ 10$^{-15}$ 
erg cm$^{-2}$ s$^{-1}$ and a luminosity upper limit of 5.0 $\times$ 10$^{37}$ erg s$^{-1}$ for 
our new SNR candidates excluding SNR1, SNR2 and SNR8. 
Our spectroscopically detected SNRs have shock velocities $<$ 85 kms$^{-1}$  indicating
that they fall into a low shock velocity range implying very old remnants from which  little X-ray emission is expected. 

According to the SN rate calculation of Matonick \& Fesen (1997) in such galaxies, 
 about half of all 
 SNe are of Type Ib/c or Type II, produced by massive stars. In turn, only half of
 these SNe are located in regions with enough ambient density
to produce a detectable SNR. This means that, only about a
quarter of all SN events may leave easily detectable optical remnants. 
In this respect four times more SN events should have exploded in M74 and we detected nine SNRs.  
With this reasoning,
we would expect to see $\sim$ 40 SNRs. If the optically observable lifetime of an SNR is assumed to be 
$\sim$ 20,000 years (Braun, Goss, \& Lyne 1989), 
 a crude SN occurrence 
rate for M74 can be 
estimated giving a value of about 1 per $\sim$ 500 years. In the light of observation of two recent SNe 
from M74, our result can be reconciled if we accept the fact that more SNe occur at places that are 
difficult to reach optically, (i.e. in gas-rich regions of the galaxy).

\begin {acknowledgements}
We thank the TUBITAK National Observatory (TUG) and the Special Astrophysical observatory (SAO) for their support for observing times and equipments. 
\end {acknowledgements}


\newpage

 \begin{table}
\begin{minipage}[t]{\columnwidth}
\caption{Characteristics of the interference filters used in our observations}             
\label{catalog}      
\centering          
\begin{tabular}{c c c c l l l }     
\hline\hline       
Name & $\lambda$ & FWHM \\
~ & Wavelength (\AA) & \AA \\    
\hline                    
[S II] & 6728 & 54 \\  
continuum  &6964 & 350 \\
H$\alpha$   & 6563 & 80 \\
continuum   &6446 & 123  \\
   
\hline                  
\end{tabular}
\end{minipage}
\end{table}  

\begin{table}
\begin{minipage}[t]{\columnwidth}
\caption{New optical SNR candidates detected in M74 in our observations.}             
\label{table:1}      
\centering          
\begin{tabular}{c c c c c l l }     
\hline\hline       
SNR Name & RA & DEC & [S II]/H$\alpha$ & I(H$\alpha$) \\
& (J2000.0)&(J2000.0) & &(erg cm$^{-2}$s$^{-1}$)\\  
\hline
SNR1 & 01:36:43.8 & +15:49:12.7 & 0.47 & 1.5E-14 \\
SNR2 & 01:36:31.5 & +15:48:50.3 & 0.46 & 6.0E-15\\
SNR3 & 01:36:46.9 & +15:45:15.2 & 0.91 & 2.9E-15 \\
SNR4 & 01:36:45.3 & +15:43:06.6 & 0.49 & 1.4E-14 \\
SNR5 & 01:36:48.9 & +15:43:34.4 & 0.58 & 2.8E-15 \\
SNR6 & 01:36:45.1 & +15:45:25.3 & 0.47 & 1.5E-14 \\
SNR7 & 01:36:45.9 & +15:44:46.1 & 0.45 & 1.6E-14\\
SNR8 & 01:36:45.9 & +15:44:27.1 & 0.46 & 1.6E-14 \\
SNR9 & 01:36:42.4 & +15:44:16.3 & 0.40 & 1.7E-14\\
\hline                  
\end{tabular}
\end{minipage}
\end{table}   

\begin{table*}
\caption{Relative line intensities and observational parameters for SNRs spectroscopically
observed in M74  }             
\label{table:2}      
\centering          
\begin{tabular}{c c c c c l l }     
\hline\hline       
Line & SNR2 & SNR3 & SNR5 \\
\hline
H$\beta$($\lambda$4861)& 100 & 100 & 100\\
OIII($\lambda$4959)&  3 &  60 &  50\\
OIII($\lambda$5007)& 27 & 122 &  160\\
NII($\lambda$5200) & - &  3 & -\\
He($\lambda$5876) &  13 & 33 & 9 \\
OI($\lambda$6300) &  17 &  21&  32\\
OI($\lambda$6364) &  14 &-&  7\\
NII($\lambda$6548) & 30 & 154&  56\\
H$\alpha$($\lambda$6563)& 288  & 589 & 335\\
NII($\lambda$6583)& 48 & 281 &  85\\
SII($\lambda$6716) &  92 & 297& 122\\
SII($\lambda$6731) & 41  & 239 &  73\\
\hline 
E$_{(B-V)}$& 0.07 & 0.06     & 0.06\\
I(H$\alpha$ )& 6.0E-15 erg cm$^{-2}$s$^{-1}$&  2.9E-15 erg cm$^{-2}$s$^{-1}$   &  2.8E-15 erg cm$^{-2}$s$^{-1}$\\
$[SII]$/H$\alpha$ & 0.46 & 0.91 &  0.58\\
\hline            
\end{tabular}
\end{table*}      
\begin{table*}
\begin{minipage}[t]{\columnwidth}
\caption{Spatial associations  between optically identified SNRs and Chandra-detected X-Ray sources (from the 
combined 2001 June - October observation).}             
\label{table:3}      
\centering          
\begin{tabular}{c c c c c l l }     
\hline\hline       
Optically detected SNR & RA & DEC & X - Ray Source \\
  
\hline
SNR1 & 01:36:43.8 & +15:49:12.7 & CXOUJ013644.0+154908 \\
SNR2 & 01:36:31.5 & +15:48:50.3 & CXOUJ013631.7+154848 \\
SNR8 & 01:36:45.9 & +15:44:27.1 & CXOUJ013646.0+154422 \\
\hline                  
\end{tabular}
\end{minipage}
\end{table*}   

\begin{figure*}
\centering
  \includegraphics[width=12cm]{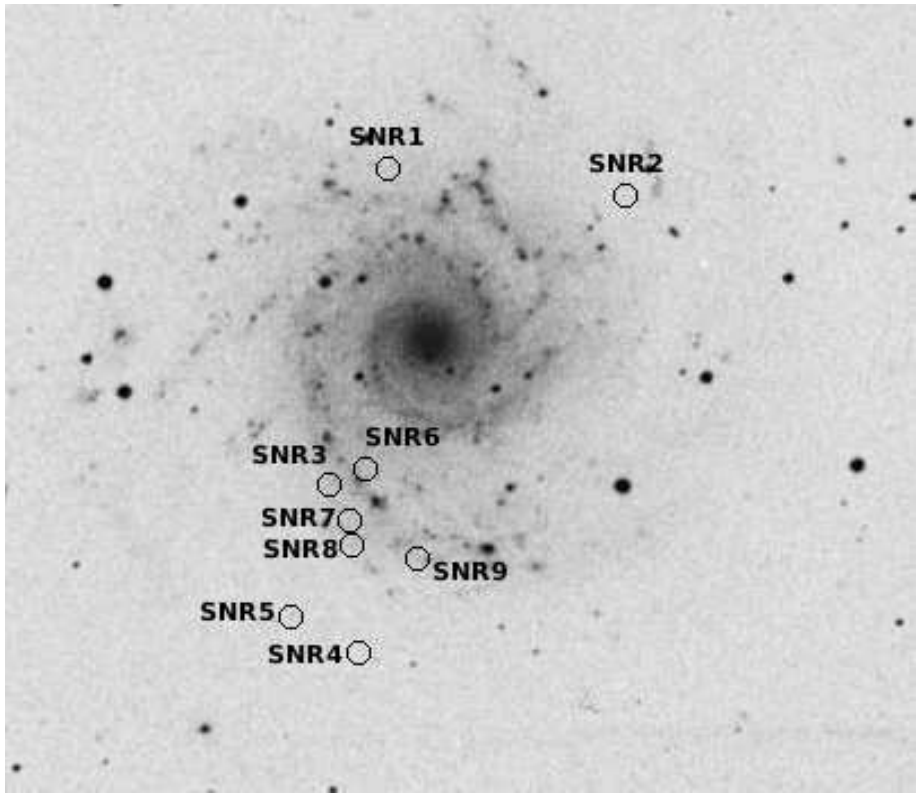}
    \caption{Nine new SNR candidates found in this work are labeled with circles on the image of M74 extracted from Digital Sky Survey (DSS).}
    \label{<Your label>}
\end{figure*}


\begin{figure*}
\centering
\subfigure[a]
{
\label{fig:sub:a}
\includegraphics[width=8cm]{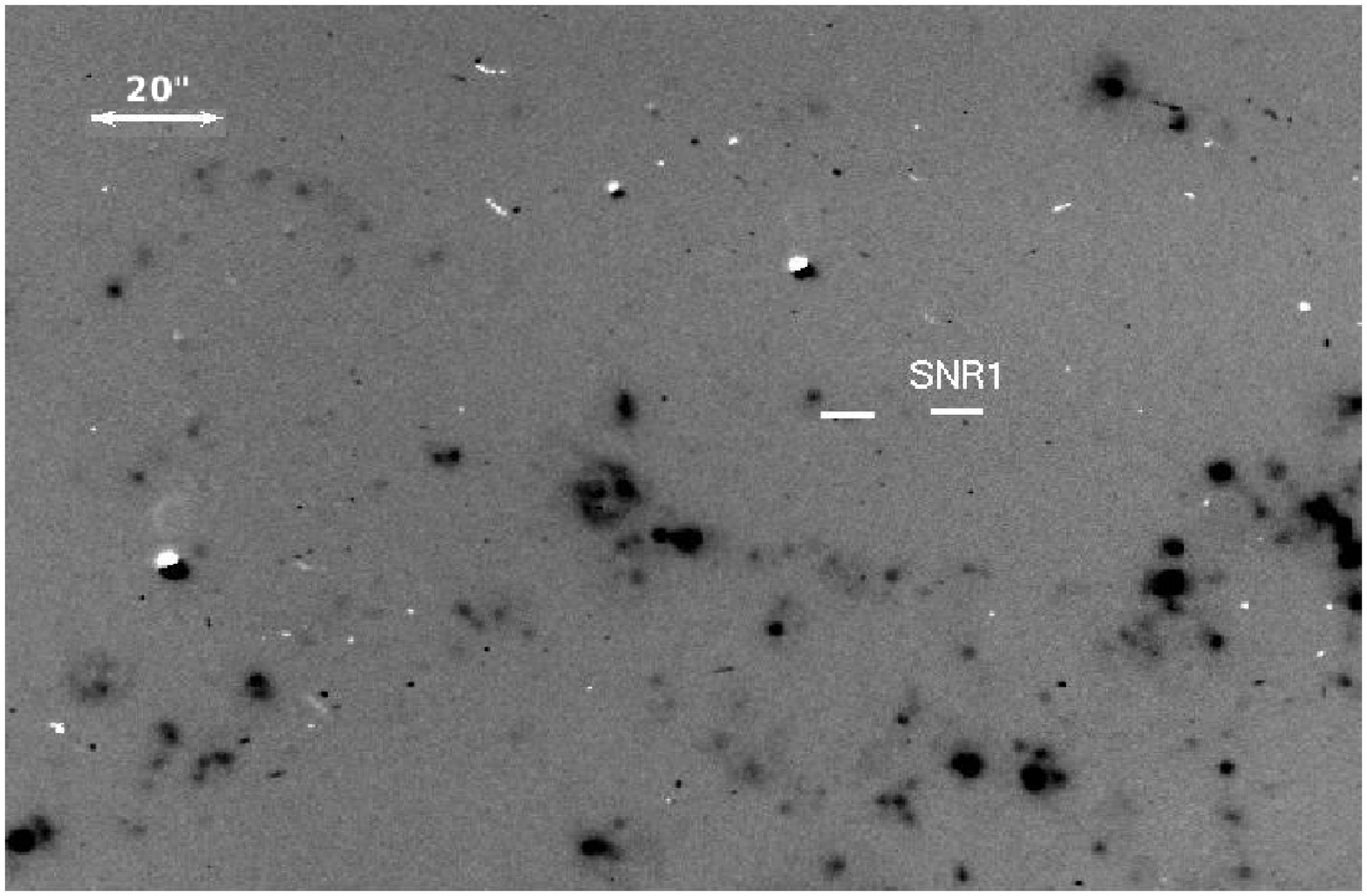}
}
\hspace{1cm}
\subfigure(b)
{
\label{fig:sub:b}
\includegraphics[width=8cm]{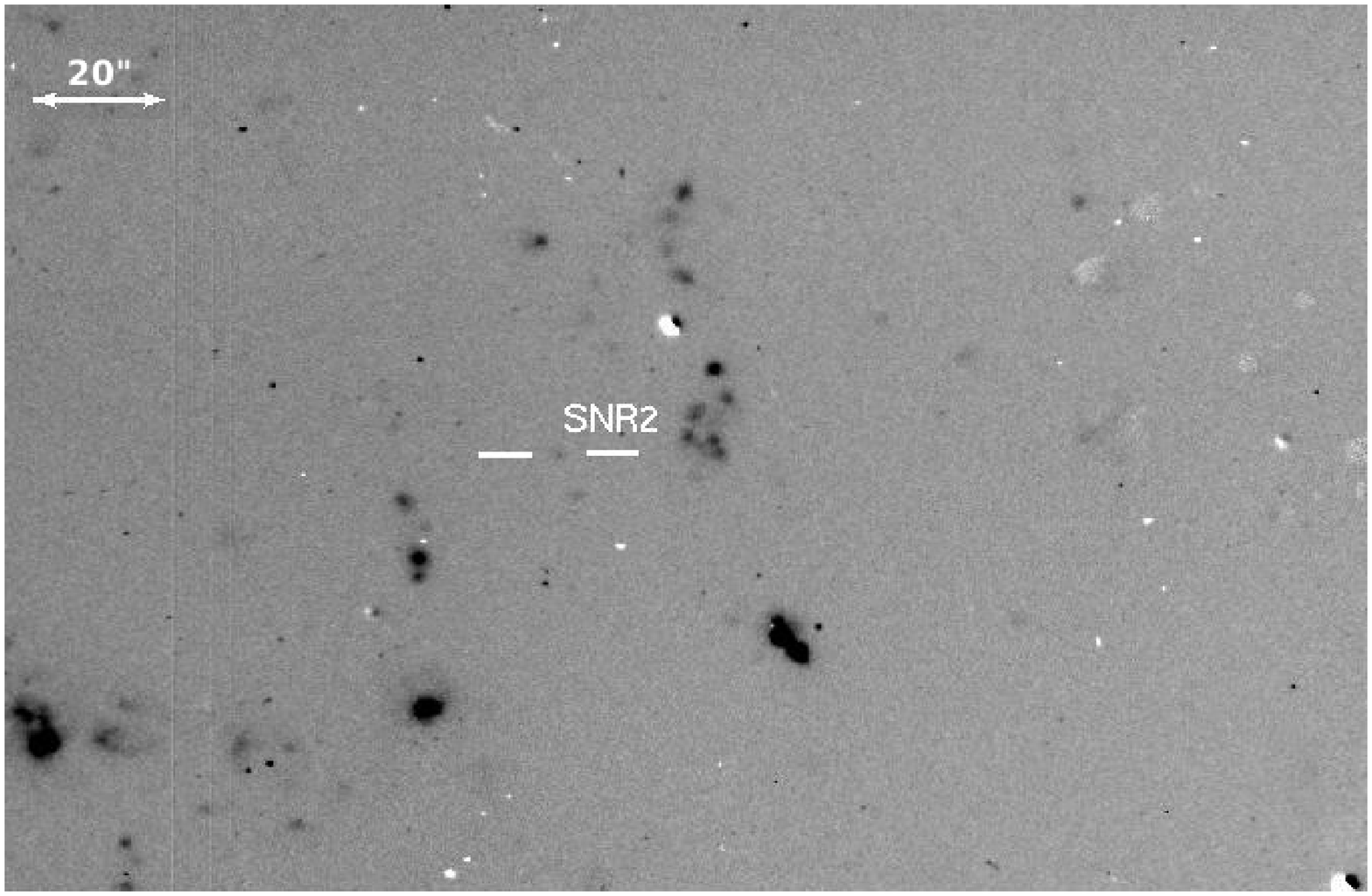}
}
\hspace{1cm}
\subfigure(c)
{
\label{fig:sub:c}
\includegraphics[width=8cm]{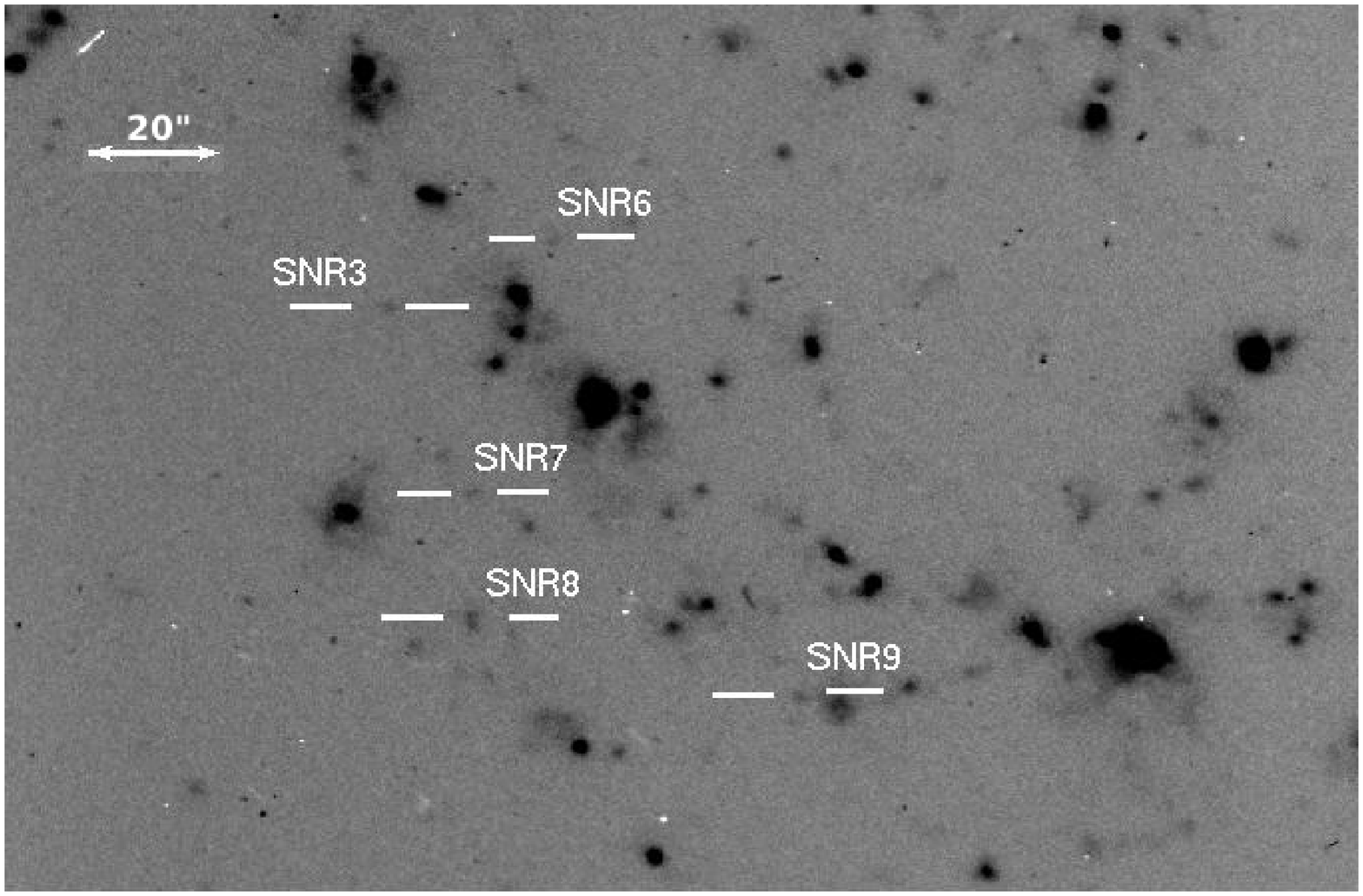}
}
\hspace{1cm}
\subfigure(d)
{
\label{fig:sub:d}
\includegraphics[height=5.25cm,width=8cm]{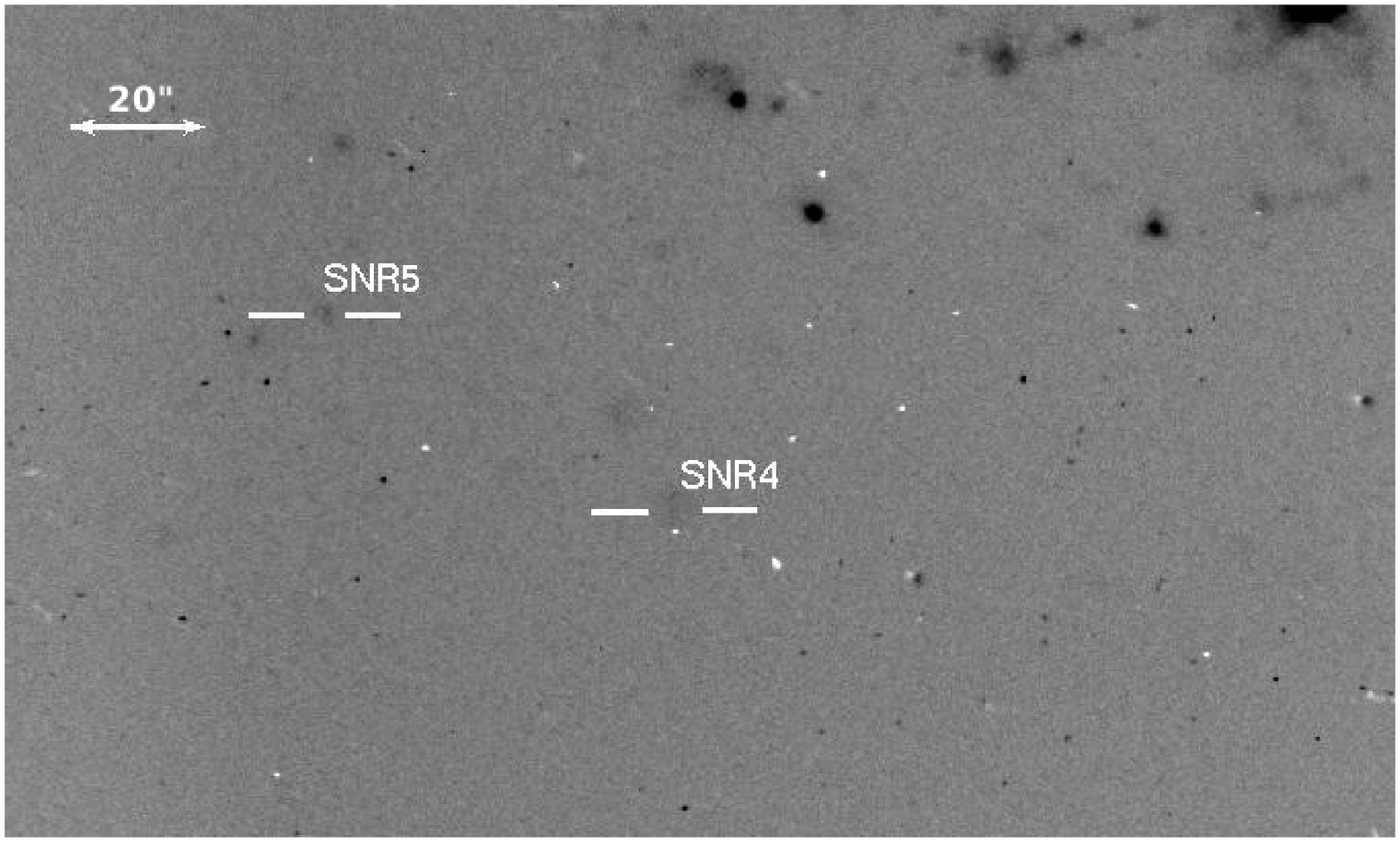}
}
\caption{ Positions of SNR1 (a), SNR2 (b),  SNR3, SNR6, SNR7, SNR8, SNR9 (c), and SNR4, SNR5 (d) overlaid on a $\sim$ 3 ${\arcmin}$.5 $\times$ 2 
${\arcmin}$.5 subfields of continuum-subtracted H$\alpha$ images of M74.}
\label{fig:sub}
\end{figure*}

  
\begin{figure}
   \centering
\includegraphics[height=6cm,width=7cm]{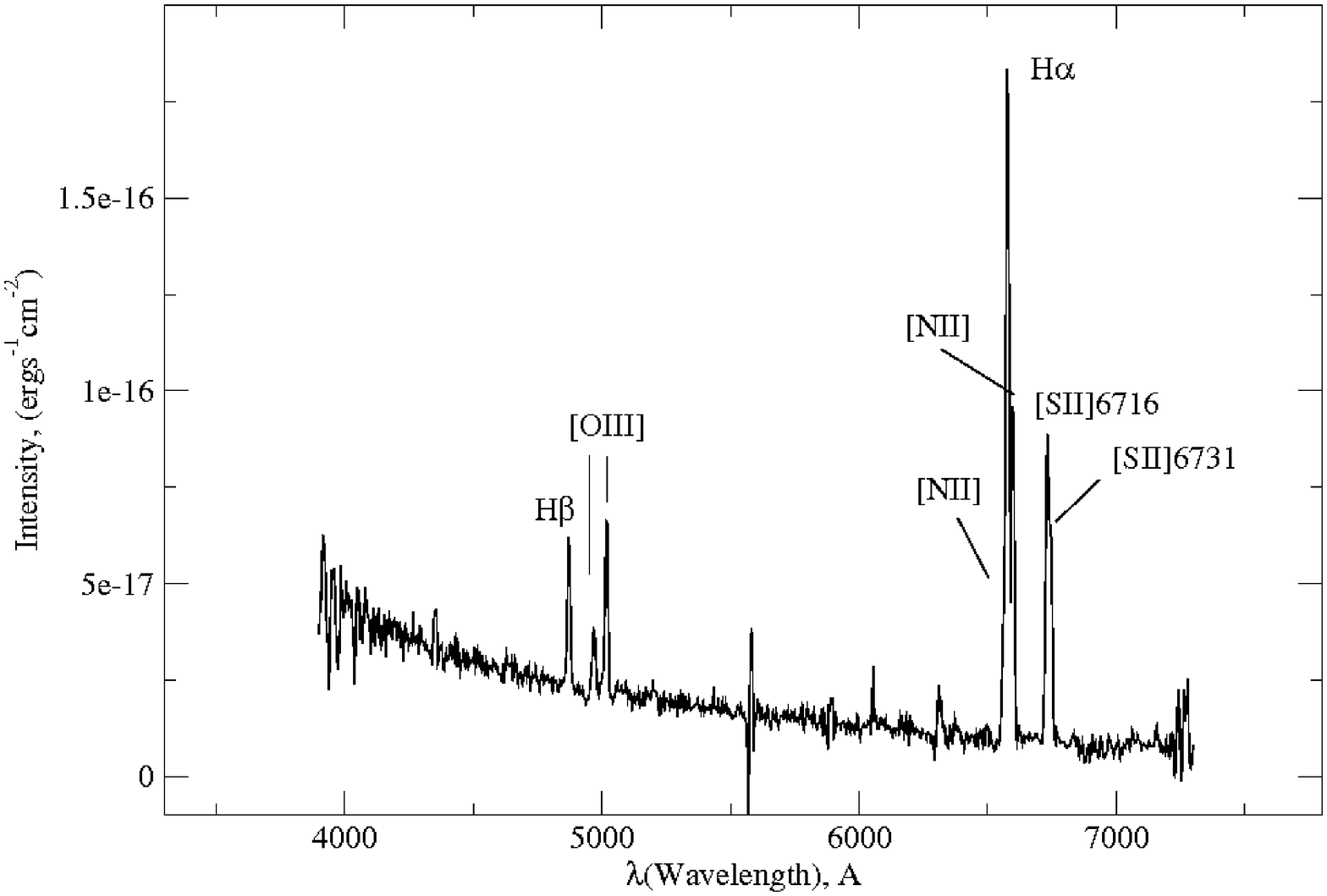}
\caption{Optical spectrum of SNR3 in M74 obtained with BTA. Identified lines are indicated.}
 \label{FigVibStab1}
   \end{figure} 
\begin{figure}
   \centering
\includegraphics[height=6cm,width=7cm]{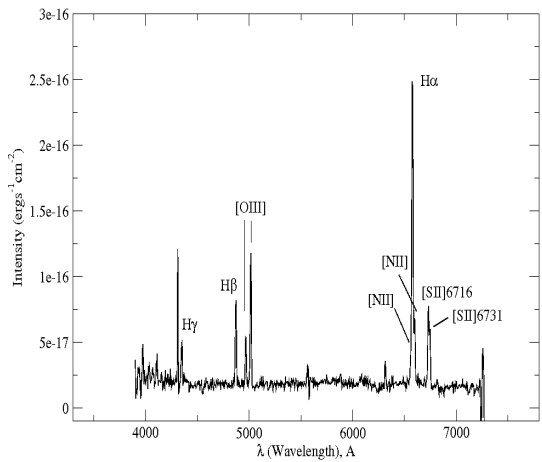}
\caption{Optical spectrum of SNR5 in M74 obtained with BTA. Identified lines are indicated. (The strong spike just blueward of H$\gamma$ is most probably due to an unsubtracted cosmic hit on the CCD chip in use.)}
 \label{FigVibStab2}
   \end{figure} 

\begin{figure}
   \centering
\includegraphics[height=6cm,width=7cm]{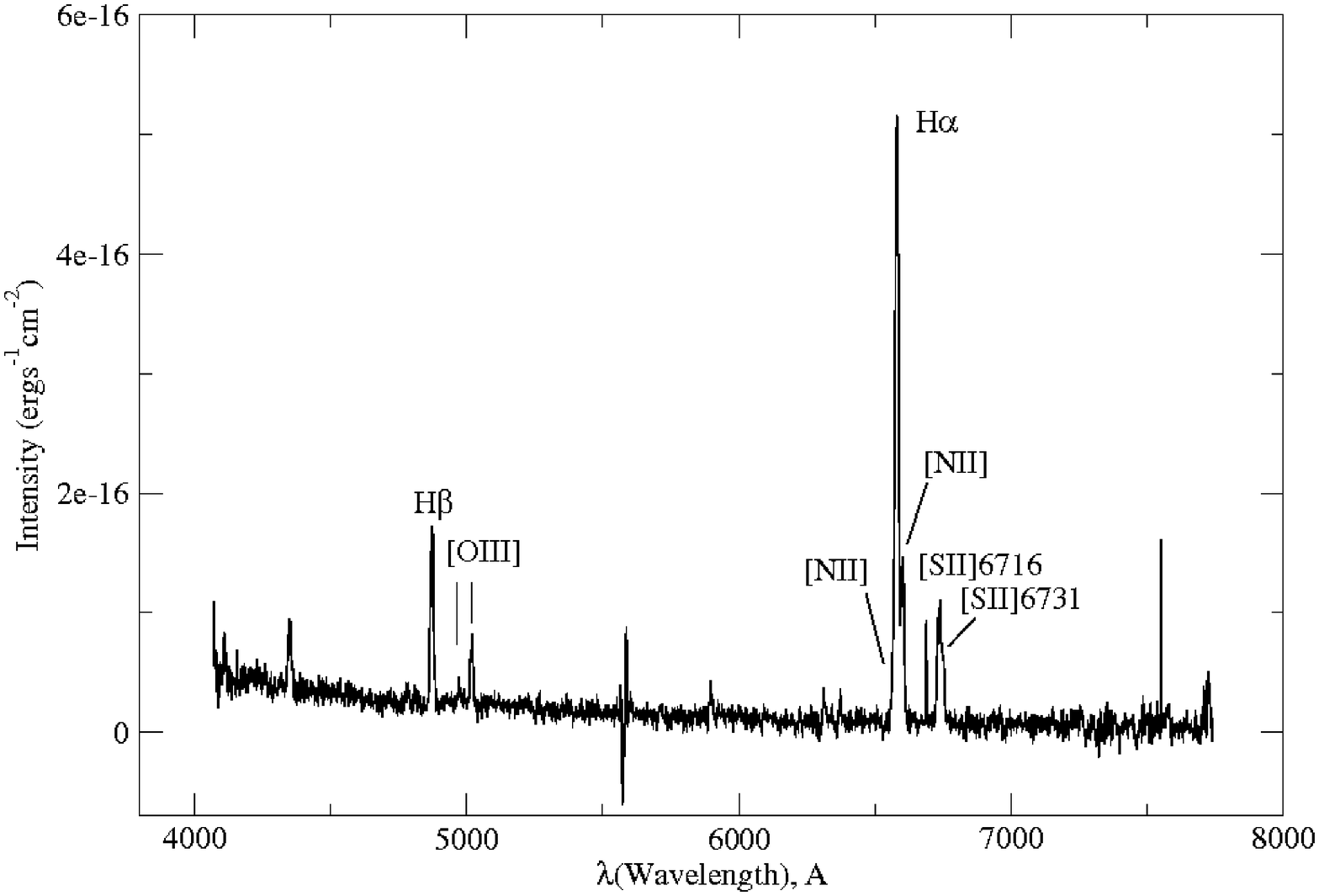}
\caption{Optical spectrum of SNR2 in M74 obtained with BTA. Identified lines are indicated.}
 \label{FigVibStab3}
   \end{figure} 


\begin{figure*}
\centering
  \includegraphics[width=12cm]{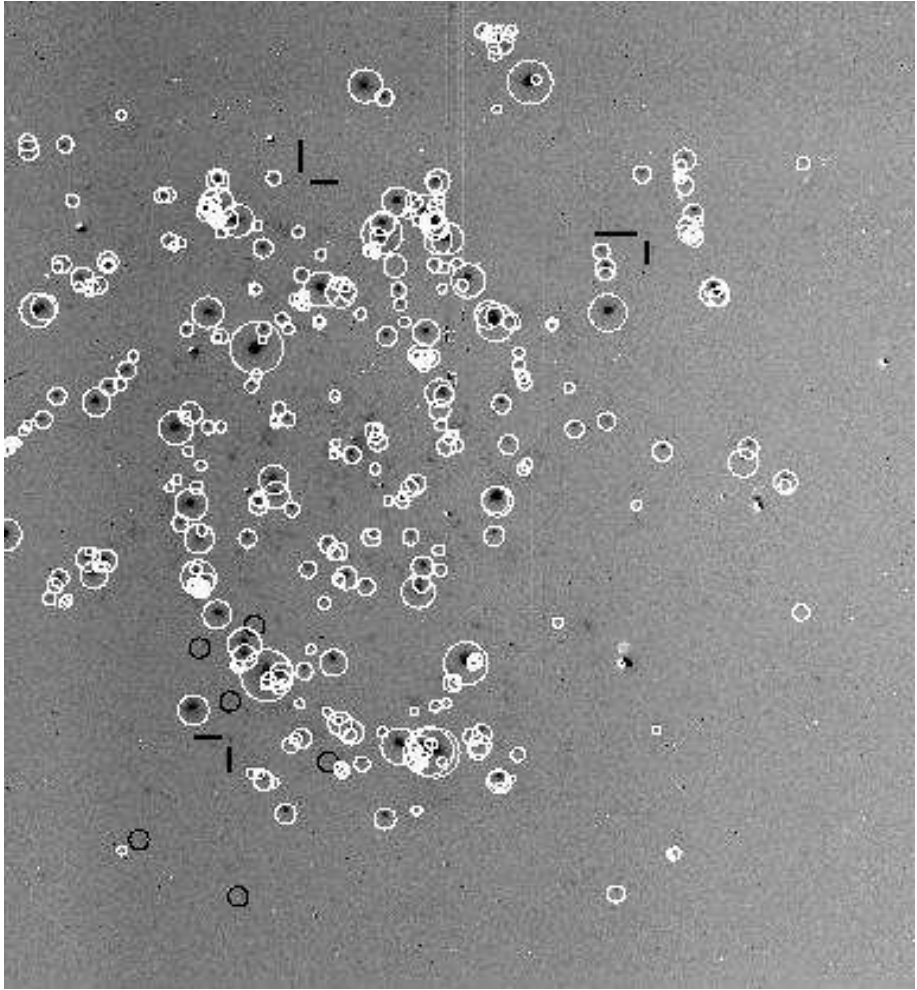}
\caption{H II regions and detected SNRs of M74. Perpendicular short lines indicate SNR1, SNR2 and SNR3 that are 
associated with Chandra X - ray point sources CXOUJ013631.7+154848, CXOUJ013644.0+154908, CXOUJ013646.0+154422, respectively. White circles indicate H II regions, and black circles indicate our six other SNRs.}
 \label{FigVibStab4}
\end{figure*}


\begin{figure*}
   \centering
\includegraphics[height=11cm,width=12cm]{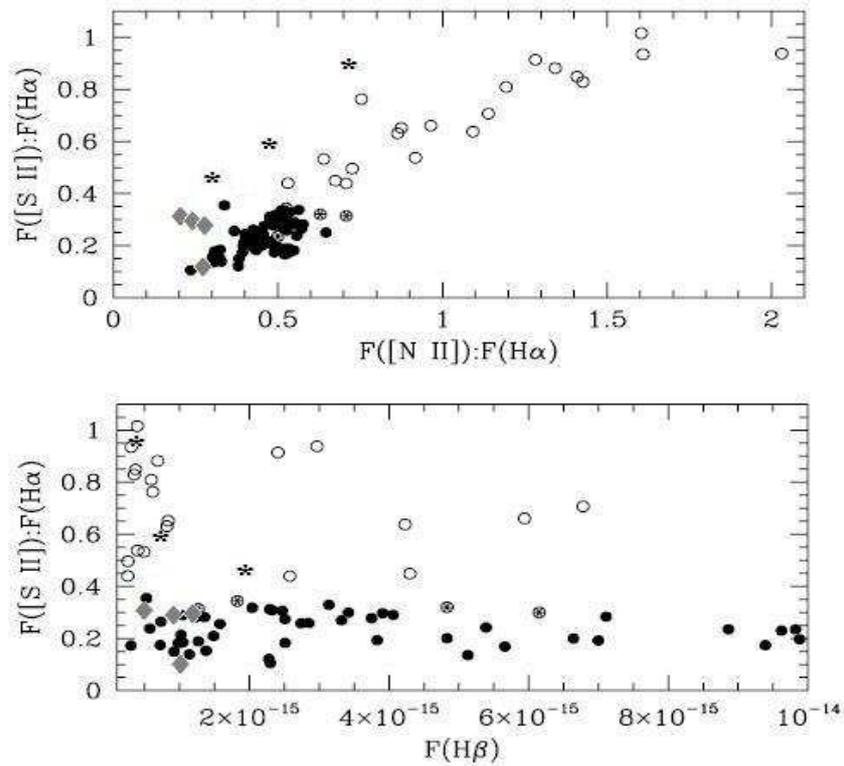}
\caption{[S II]/H$\alpha$ vs [NII]/H$\alpha$ ratios and [S II]/H$\alpha$ vs H$\beta$ fluxes. Stars indicate our spectroscopically identified SNRs while diamonds those unidentified ones. These are compared with SNRs of M83 Blair \& Long (2004) where open circles refer to spectroscopically identified SNRs, semifilled circles refer to spectroscopically unidentified SNRs, and filled circles refer to H II regions. Similar tendencies between SNRs of M83 and M74 are noticeable.}
\label{fign}
\end{figure*}

\begin{figure}
   \centering
\includegraphics[height=8cm,width=6.8cm,angle=-90]{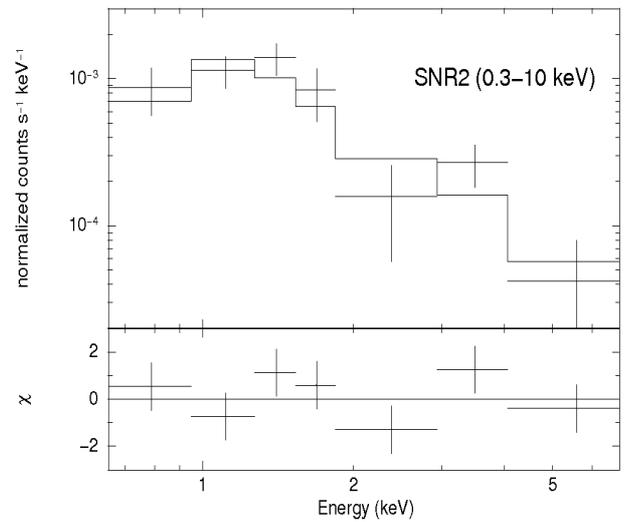}
\caption{X-ray spectrum of SNR2 obtained using archival Chandra data in 0.3-10.0 keV fitted with
a neutral hydrogen absorption model, WABS and a PSHOCK non-equilibrium ionization
plasma emission model ({\tt WABS*PHSOCK}). The crosses indicate the data, the solid line is the fitted model(upper panel) and 
 the residuals between the data and the
model in standard deviations (lower panel).}
 \label{Figxray}
   \end{figure}
\end{document}